\documentclass[amsmath,amssymb,superscriptaddress,nobalancelastpage,prb,twocolumn,longbibliography]{revtex4-1}

\usepackage{graphicx}
\usepackage{varioref}
\usepackage{xr-hyper}
\usepackage{xcolor}
\usepackage{nicefrac}
\usepackage{xfrac}
\usepackage{hyperref}
\hypersetup{colorlinks,linkcolor=blue,urlcolor=blue,citecolor=orange}
\usepackage{ulem}
\usepackage{siunitx}
\usepackage{graphicx}
\usepackage{dcolumn}
\usepackage{bm}
\usepackage{braket}
\usepackage{wasysym}
\usepackage{textcomp}

\newcommand{\lsco}{{La}$_{1.88}${Sr}$_{0.12}${CuO}$_4$}

\begin{document}

\title{Uniaxial-pressure-induced Stripe-order in a High-temperature Superconductor}

\title{Strain-Stabilized Stripe-order in a High-temperature Superconductor}

\title{Strain-Induced Checkerboard to Stripe-order Transition in a Cuprate Superconductor}

\title{Strain-Induced Uniaxial Stripe-order in a Cuprate Superconductor}

\title{ Stripe-Order-Degeneracy Squeeze-Out in  a Cuprate Superconductor}

\title{ Lifting Quantum Domain Degeneracy in a Cuprate Superconductor}

\title{ Lifting Domain Degeneracy of an Intertwined Quantum Matter State}

\title{Disentangling Intertwined Quantum States in a Prototypical High-temperature Superconductor}

\title{Disentangling Intertwined Quantum States in a Prototypical Cuprate Superconductor}

\author{J. Choi}
\affiliation{Physik-Institut, Universit\"{a}t Z\"{u}rich, Winterthurerstrasse 190, CH-8057 Z\"{u}rich, Switzerland}

\author{Q. Wang}
\affiliation{Physik-Institut, Universit\"{a}t Z\"{u}rich, Winterthurerstrasse 190, CH-8057 Z\"{u}rich, Switzerland}

\author{S. J\"{o}hr}
\affiliation{Physik-Institut, Universit\"{a}t Z\"{u}rich, Winterthurerstrasse 190, CH-8057 Z\"{u}rich, Switzerland}

\author{N.~B. Christensen}
\affiliation{Department of Physics, Technical University of Denmark, DK-2800 Kongens Lyngby, Denmark}

\author{J. K\"{u}spert}
\affiliation{Physik-Institut, Universit\"{a}t Z\"{u}rich, Winterthurerstrasse 190, CH-8057 Z\"{u}rich, Switzerland}

\author{D. Bucher}
\affiliation{Physik-Institut, Universit\"{a}t Z\"{u}rich, Winterthurerstrasse 190, CH-8057 Z\"{u}rich, Switzerland}

\author{D. Biscette}
\affiliation{Physik-Institut, Universit\"{a}t Z\"{u}rich, Winterthurerstrasse 190, CH-8057 Z\"{u}rich, Switzerland}

\author{M. H\"{u}cker}
\affiliation{Department of Condensed Matter Physics, Weizmann Institute of Science, Rehovot 7610001, Israel}

\author{T.~Kurosawa}
\affiliation{Department of Physics, Hokkaido University - Sapporo 060-0810, 
Japan}
 
\author{N.~Momono}
\affiliation{Department of Physics, Hokkaido University - Sapporo 060-0810, 
Japan}
\affiliation{Department of Applied Sciences, Muroran Institute of Technology, 
Muroran 050-8585, Japan}

\author{M.~Oda}
\affiliation{Department of Physics, Hokkaido University - Sapporo 060-0810, 
Japan}

\author{O. Ivashko}
\email{oleh.ivashko@desy.de}
\affiliation{Deutsches Elektronen-Synchrotron DESY, Notkestra{\ss}e 85, 22607 Hamburg, Germany.}

\author{M.~v.~Zimmermann}
\affiliation{Deutsches Elektronen-Synchrotron DESY, Notkestra{\ss}e 85, 22607 Hamburg, Germany.}

\author{M. Janoschek}
\affiliation{Physik-Institut, Universit\"{a}t Z\"{u}rich, Winterthurerstrasse 190, CH-8057 Z\"{u}rich, Switzerland}
\affiliation{Laboratory for Neutron and Muon Instrumentation, Paul Scherrer Institut, CH-5232 Villigen PSI, Switzerland}

\author{J. Chang}
\email{johan.chang@physik.uzh.ch}
\affiliation{Physik-Institut, Universit\"{a}t Z\"{u}rich, Winterthurerstrasse 190, CH-8057 Z\"{u}rich, Switzerland}

\maketitle
\textbf{Spontaneous symmetry breaking constitutes a paradigmatic classification scheme of matter. However, broken symmetry also entails domain degeneracy that often impedes identification of novel low symmetry states. In quantum matter, this is additionally complicated 
by competing intertwined symmetry breaking orders. A prime example is that of unconventional superconductivity and density-wave orders in doped cuprates in which their respective symmetry relation remains a key question. Using uniaxial pressure as a domain-selective stimulus in combination with x-ray diffraction, we unambiguously reveal that the fundamental symmetry of the charge order in the prototypical cuprate La$_{1.88}$Sr$_{0.12}$CuO$_4$ is characterized by uniaxial stripes. We further demonstrate the direct competition of this stripe order with unconventional superconductivity via magnetic field tuning.  The stripy nature of the charge-density-wave state established by our study is a prerequisite for the existence of a superconducting pair-density-wave -- a theoretical proposal that clarifies the interrelation of intertwined quantum phases in unconventional superconductors -- and paves the way for its high-temperature realization.}\\[1mm]

From subatomic to cosmological length scales, spontaneously broken symmetry represents a conceptual cornerstone with broad relevance in fields such as physics, chemistry, biology and medicine. In condensed matter research, symmetry allows for the classification of matter and can often provide critical insights in their  properties without knowledge of microscopic details~\cite{SondhiRMP1997,VojtaRPP2003}. Notably, in quantum materials, which exhibit a delicate interplay between multiple nearly-degenerate states of matter, symmetry is key to understanding the electronic properties. However, experimentally, this inherent complexity of multiple coupled phases is a major challenge in determining the all-important symmetry~\cite{SymmetryNatPhys}. This is exemplified by a plethora of quantum phases for which the symmetry is not known such as several ``hidden order" states~\cite{Mydosh}, or exotic superconducting states with complex symmetry~\cite{NatureUTe2,NatureSr2RuO4}.

\begin{figure}[t]
\center{\includegraphics[width=0.48\textwidth]{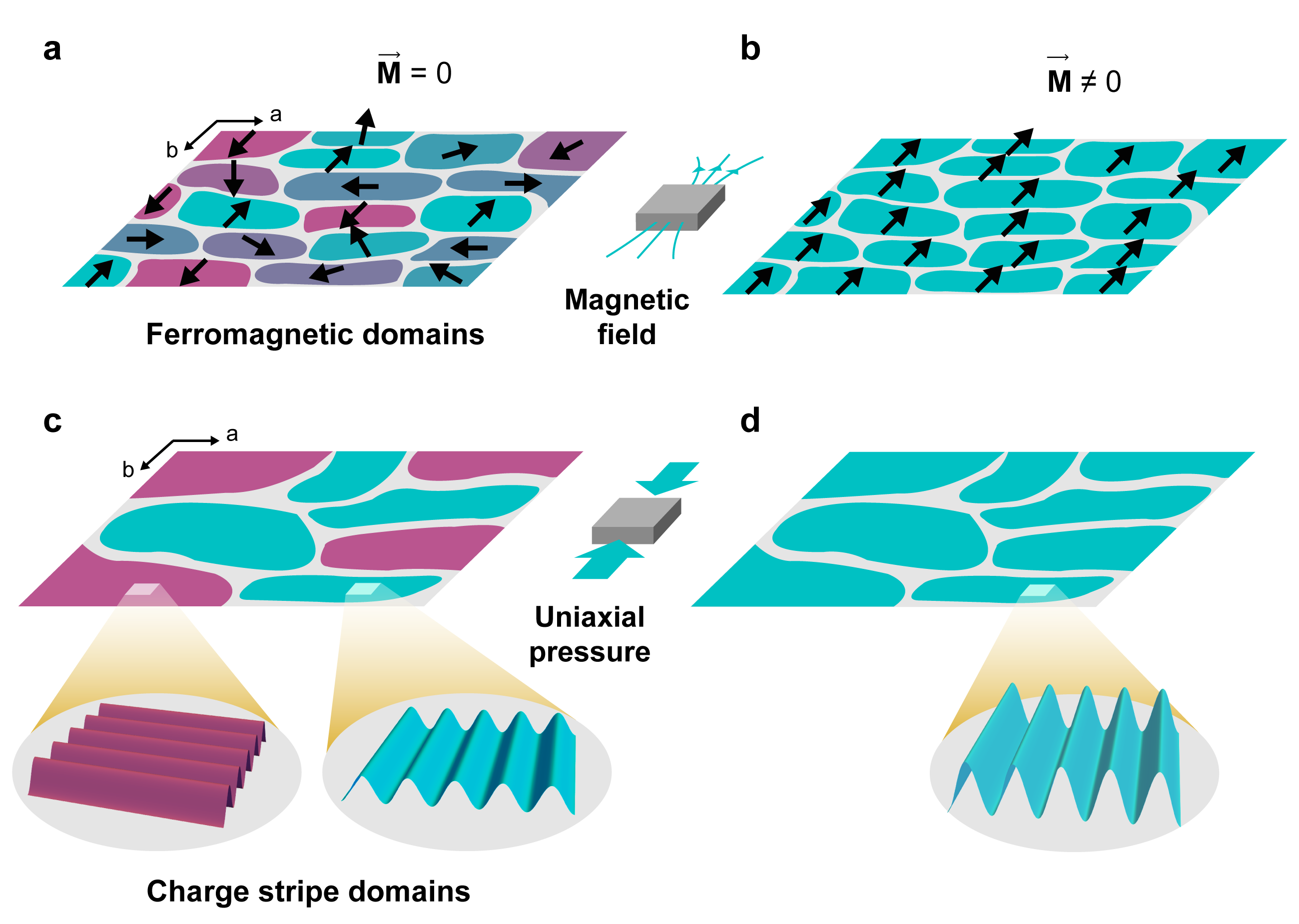}} 	
\caption{\textbf{Lifting of domain degeneracy in quantum matter.} Symmetry breaking is often associated with domain formation. (\textbf{a}) Across a ferromagnetic transition, spin rotational symmetry is broken with respect to the underlying atomic lattice. For example, domains with an arbitrary direction of magnetization form in the absence of external stimulus. (\textbf{b}) This domain degeneracy can be lifted by application of an external magnetic field. (\textbf{c,d}) Analogously, by applying  uniaxial strain, the degeneracy of charge stripe order is removed in the cuprate superconductor La$_{1.88}$Sr$_{0.12}$CuO$_4$.}
\label{fig1}
\end{figure} 

\begin{figure*}[t]
\center{\includegraphics[width=0.95\textwidth]{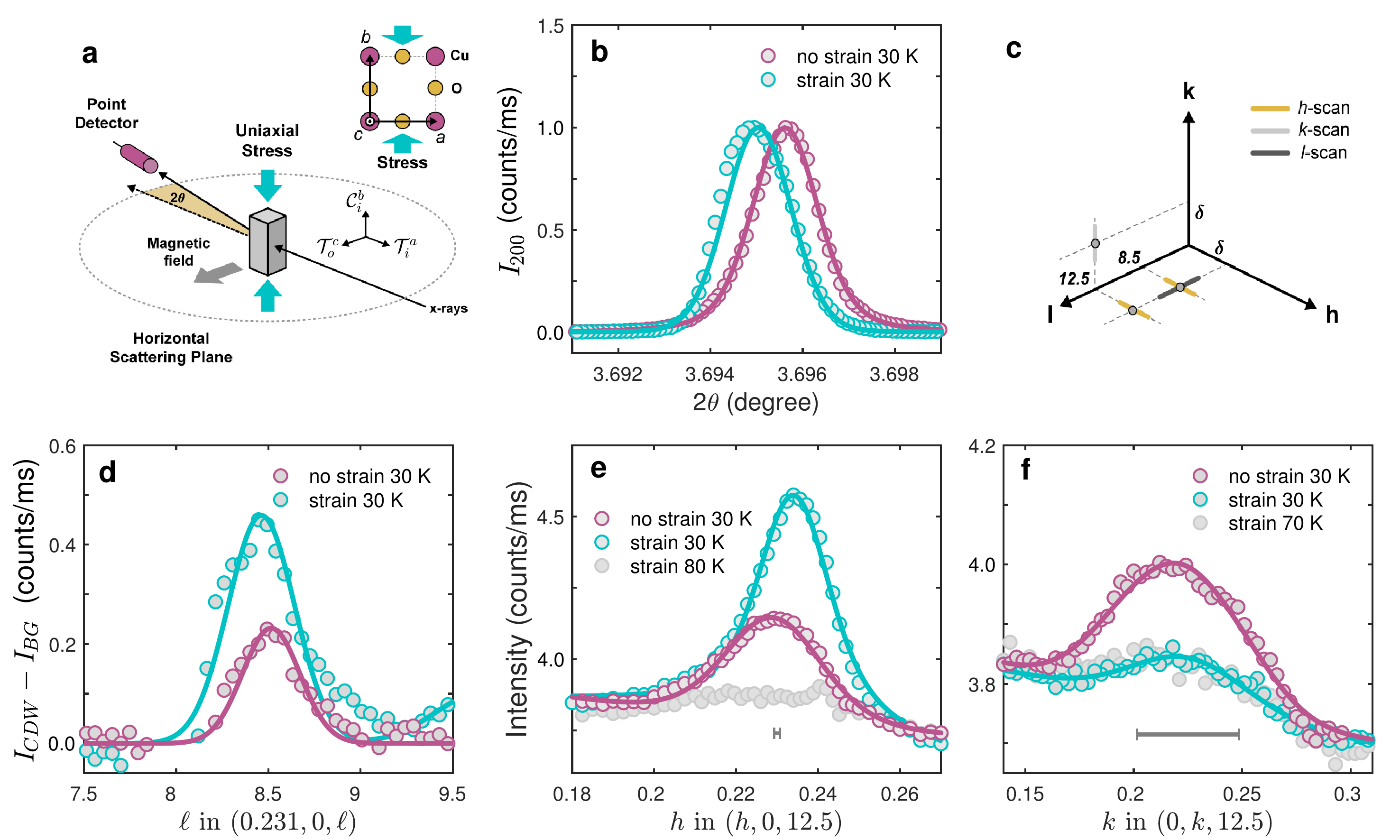}} 
\caption{ \textbf{Effect of uniaxial pressure on the charge density wave in \lsco.} (\textbf{a}) Schematic of the experimental scattering geometry. Uniaxial pressure and magnetic field are applied along the $b$- and $c$-axes, respectively. (\textbf{b}) Demonstration of how uniaxial pressure along the $b$-axis enhances the $a$-axis lattice parameter -- smaller scattering angle $2\theta$ implies larger Cu-O-Cu distance. 
(\textbf{c}) Schematic illustration of the studied part of reciprocal space. Yellow, grey and dark grey lines indicate respectively $h$-, $k$-, and $\ell$-scans through CDW reflections. Panels (\textbf{d-f}) show diffraction intensity of $h$-, $k$-, and $\ell$- scans through charge-density-wave ordering vectors without (purple) and with (cyan) application of uniaxial strain, respectively. For the $\ell$-scan (\textbf{d}), covering two Brillouin zones (BZs), background subtraction was performed. The $h$- and $k$-scans cover only a small fraction of a BZ and hence no subtraction is required.  
Error bars in (\textbf{b,d-f}), set by counting statistics, are
smaller than the used symbols.
Horizontal bars in (\textbf{e,f}) indicate the instrumental momentum resolution and solid lines are fits to a Voigt profile.}
\label{fig2}
\end{figure*} 

Unconventional superconductivity and symmetry breaking phases are often found to coexist. In cuprate materials, high-temperature superconductivity microscopically intertwines with a multitude of electronic quantum states~\cite{FradkinRMP2015} including a mysterious pseudogap phase~\cite{NormanAP2005}, charge-density-wave (CDW) order~\cite{tranquada,Wu11}, as well as electronic nematic phases~\cite{MurayamaNC2019,DaouNat2010}. The mechanism of superconductivity, the nature of the pseudogap phase and the symmetry properties of the density-wave states remain to be clarified. A central question is whether the entire problem derives from a primary order (a mother state) that breaks all relevant symmetries and induces the formation of descendant phases.

A pair-density-wave (PDW) — a superconducting state characterized by an order parameter that is spatially modulated in such a way that its spatial average vanishes — is a major contender for such a “mother state”~\cite{AgterbergAnnRev2020}. Determination of the CDW order symmetry is an imperative step in the search for the putative pair-density-wave order. Experimentally, the outstanding challenge is that due to the existence of domains, the diffraction signatures of the proposed uniaxial stripe CDW order is virtually indistinguishable from biaxial structures in which the charge density is simultaneously modulated along two perpendicular directions.

We address this experimental constraint by employing a domain-selective stimulus to unambiguously establish the existence of unidirectional 
CDWs in cuprates. This is illustrated in Fig.~\ref{fig1} for the example of spontaneously broken rotational symmetry in a ferromagnet. Notably, spontaneous ferromagnetism involves the formation of domains, leading to a vanishing net magnetization.
Only the presence of a weak external magnetic field lifts the domain degeneracy and reveals the low-symmetry properties. 

\begin{figure*}[t]
\center{\includegraphics[width=0.85\textwidth]{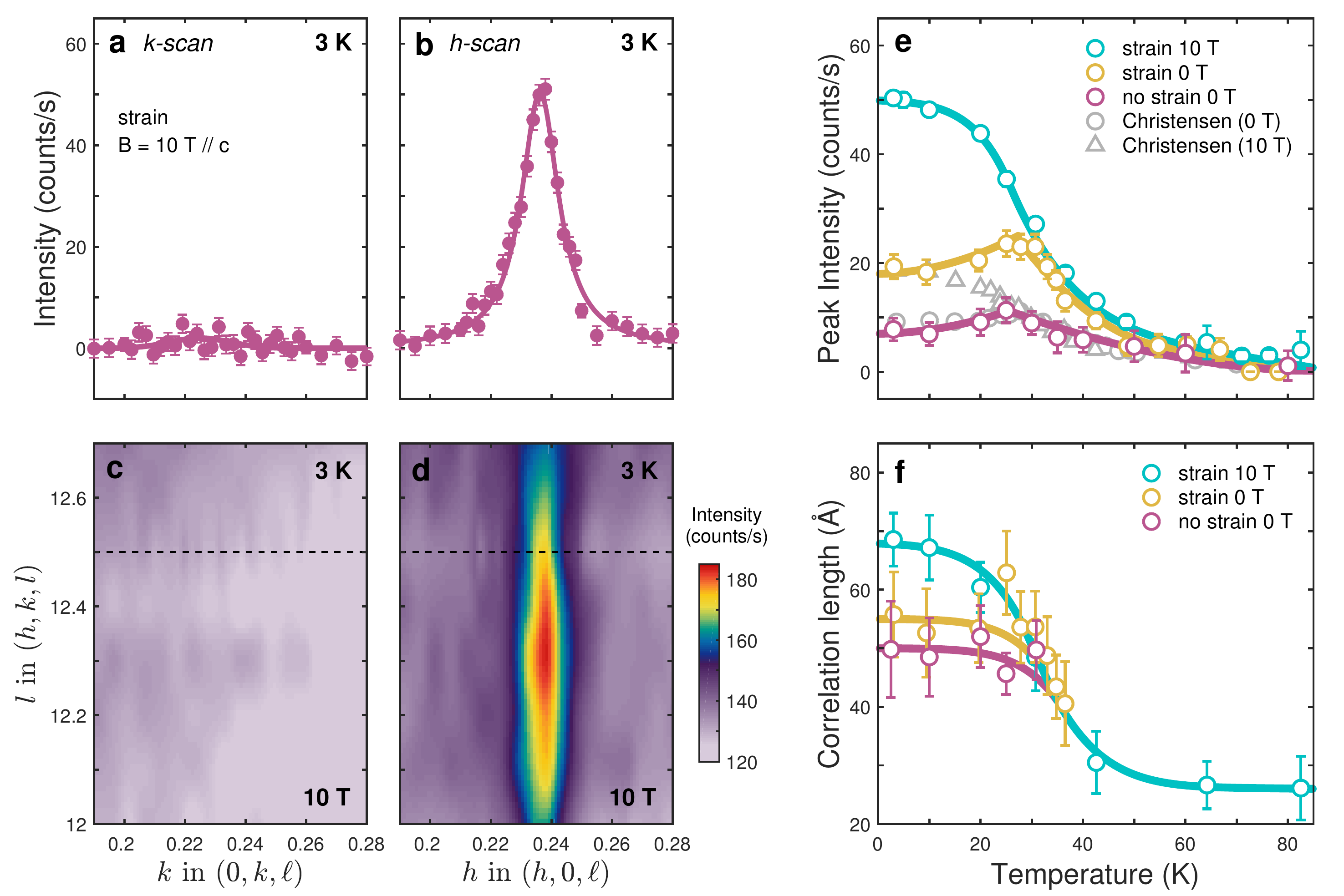}} 	
\caption{\textbf{Magnetic field and temperature dependence of strain-induced charge stripe order.} (\textbf{a,b}) Momentum $h$- and $k$-scans recorded in a $c$-axis magnetic field of 10 T and with compressive strain along the $b$-axis direction for $\ell=12.5$. Absence of charge order along the 
direction of applied strain reveals the unidirectional 
nature of the charge density wave. Error bars are set by counting statistics. (\textbf{c,d}) Diffraction intensity, displayed in false color scale, around respectively $(0,\delta,12.5)$  and $(\delta,0,12.5)$ recorded with temperature and magnetic field as indicated. (\textbf{e,f}) CDW peak amplitude and correlation length as a function of temperature for magnetic field and strain conditions as indicated. Grey data points are adopted from  Ref.~[\onlinecite{Christensen14}]. Error bars reflect standard deviations of the fits (See Methods section).}
\label{fig3}
\end{figure*} 

Our key finding is that application of uniaxial pressure
applied to \lsco\ (LSCO) lifts domain degeneracy (see Fig.~\ref{fig1}) and thereby uncovers a unidirectional stripe structure. The in-plane charge stripe ordering vector is perpendicular to the uniaxial stress. Application of magnetic field amplifies the stripe order within the superconducting state, demonstrating a unidirectional phase competition. 
Being generated by a modest uniaxial pressure stimulus to La$_{2-x}$Sr$_x$CuO$_4$, the stripe order emerges as an intrinsic electronic property.

Charge-density-wave  order manifests itself, in La-based cuprates, by satellite reflections at $\mathbf{Q}=\bm{\tau}+\mathbf{q}_i$. Here $\bm{\tau}$ represents fundamental lattice Bragg peaks and $\mathbf{q}_1=(\delta,0,0.5)$ and $\mathbf{q}_2=(0,\delta,0.5)$ with 
$\delta\approx 1/4$ reciprocal lattice units \cite{tranquada,HuckerPRB2011,CroftPRB2014,ThampyPRB2014} using lattice parameters $a\approx 3.8\geq b$ and $c\approx 13.2$~\AA. In-plane pressure along the $b$-axis produces a compressive ($\mathcal{C}_i^b$) strain. In turn, tensile strains propagate along the orthogonal  in- ($\mathcal{T}_i^a$) and out-of- plane ($\mathcal{T}_o^c$) directions (Fig.~\ref{fig2}a,b). Along these high-symmetry crystallographic axes, compressive or tensile strain is defined as $\epsilon_j=(u_j-u_j^0)/u_j^0$ where $u_j$ is the lattice parameter along one of the three directions
$\mathcal{C}_i^b$, $\mathcal{T}_i^a$  or $\mathcal{T}_o^c$, and $u_j^0$ refers to the zero-pressure lattice constants.

Starting in the normal state ($T=30$~K), uniaxial $b$-axis pressure induces an approximately twofold enhancement of the charge order reflections found at ($\delta,0,\ell$) with $\ell=8.5,12.5$ (See Fig.~\ref{fig2}c-e). Along the $\mathcal{C}_i^b$ direction, by contrast, the charge order reflection $(0,\delta,12.5)$ is completely suppressed (Fig.~\ref{fig2}f). 
The uniaxial-pressure-enhanced CDW order along the $\mathcal{T}_i^a$ axis displays a temperature dependence that within a twofold scaling factor is identical to that found under ambient pressure (See Fig.~\ref{fig3}e and Extended Data Fig.~\ref{figs6}). Upon cooling into the superconducting state ($T<T_c$), the charge order is partially suppressed, as commonly found in the cuprates\cite{Hayward14}.

We find that a $c$-axis magnetic field has no impact along the $\mathcal{C}_i^b$ direction. The CDW order remains completely suppressed even in a 10 T magnetic field (Fig.~\ref{fig3}a,c). By contrast, along the $\mathcal{T}_i^a$ direction the CDW diffraction amplitude displays a strong magnetic field effect inside the superconducting state. The stripe order peak is enhanced by an another factor of $\sim$2.5 upon application of 10 T (Fig.~\ref{fig3}b,d). The in-plane correlation length ($\mathcal{T}_i^a$ direction) is only marginally improved with the application of strain. Application of magnetic field induces a more significant increase of the correlation length that 
reaches $\xi_a=70$~\AA~(Fig.~\ref{fig3}f). Neither uniaxial pressure nor magnetic field influences the out-of-plane correlation length (See Extended Data Fig.~\ref{figs7}).

\begin{figure}[t]
\center{\includegraphics[width=0.45\textwidth]{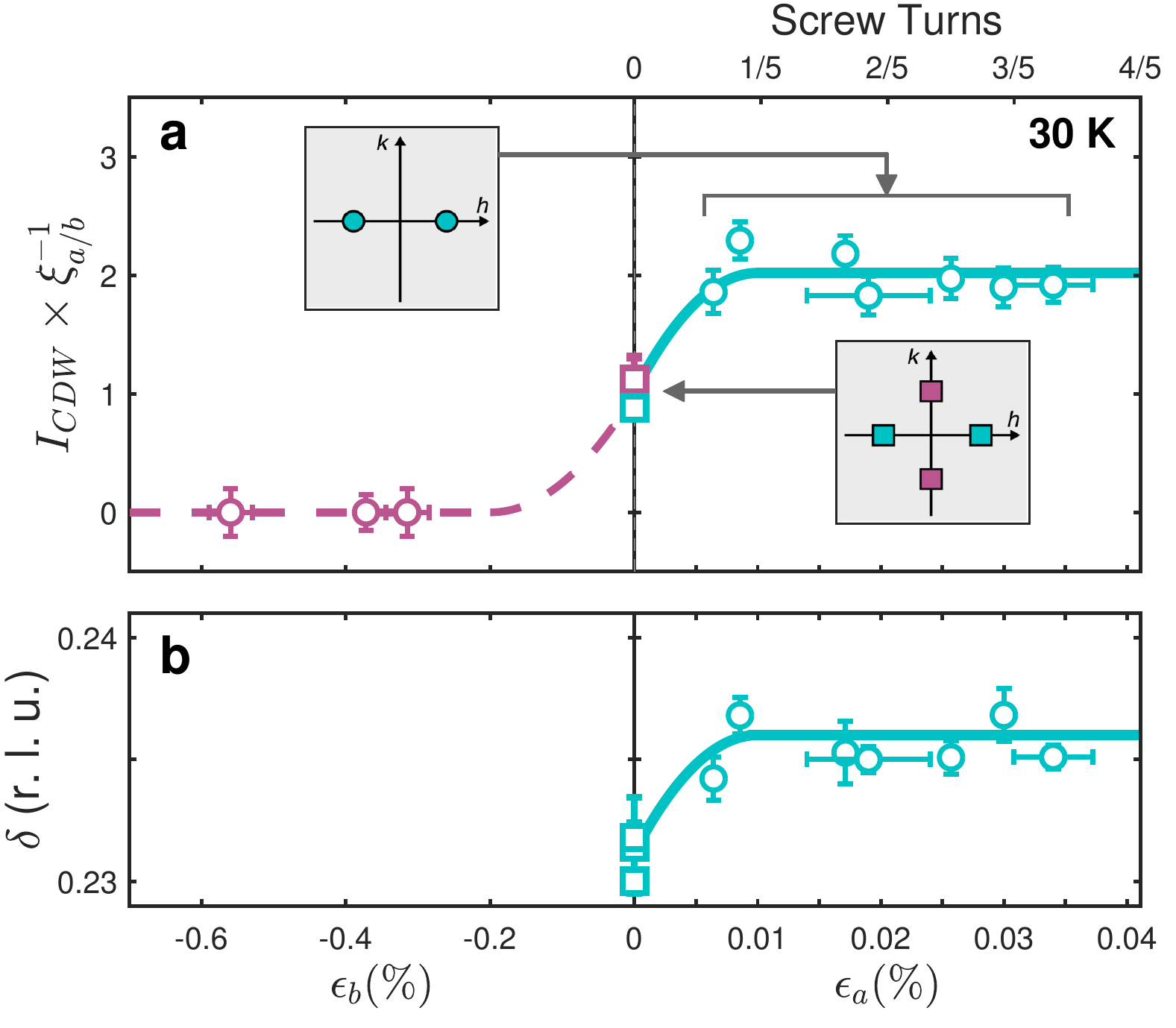}} 	
\caption{\textbf{Strain-induced stripe charge order detwinning.} (\textbf{a}) Integrated charge-density-wave intensity as a function of strain $\epsilon_a=\Delta a/a$ and $\epsilon_b=\Delta b/b$ along respective the $a-$ (tensile strain $\mathcal{T}_i^{a}$ direction) and $b-$ (compressive strain $\mathcal{C}_i^{b}$ direction) axis.  Insets illustrate the observed diffraction patterns consistent with stripe order (left) and
biaxial or 
domains of stripe order (right) respectively. (\textbf{b}) Charge density wave incommensurability, $\delta$, as a function of strain. }
\label{fig4}
\end{figure} 

The enhanced charge order diffraction intensity is insensitive to the applied stress magnitude. We find that a strain of $\epsilon_a=0.01$\% is enough to trigger the stripe order structure (Fig.~\ref{fig4}a). The twofold enhancement remains up to the largest applied strain $\epsilon_a\sim 0.03$\%. Finally, we find that the incommensurability $\delta$ is marginally larger in the stripe ordered phase (Fig.~\ref{fig4}b). 
 
This incommensurability effect could indicate a uniaxial-pressure-induced change of orthorhombic structure~\cite{RobertsonPRB2006}. Uniaxial pressure is also likely to influence the low-energy electronic structure. The band structure of LSCO has a van Hove singularity in the vicinity of the Fermi level \cite{HorioPRL2018}. Uniaxial pressure acts to push this singularity closer to the Fermi level (and eventually across) along the $\mathcal{T}_i
^b$-direction~\cite{YamasePRB2006,BarberPRL2018}. 
This increase of density of states at the Fermi level may be involved in the stripe order domain lifting. 

On a phenomenological level, charge-density modulation $\delta \rho (r)$ structures are described by~\cite{RobertsonPRB2006,MaestroPRB2006}:

\begin{equation}\nonumber
    \delta \rho (r)=\textrm{Re}(\Phi_x e^{i\bf{q_1}\cdot\bf{r}}) + \textrm{Re}(\Phi_y e^{i\bf{q_2}\cdot\bf{r}})
\end{equation}
where $\bf{r}$ is a two-dimensional spatial coordinate. The amplitudes $\vert \Phi_x\vert$ and $\vert \Phi_y\vert$ are non-zero for biaxial structures. Stripe order, by contrast, refers to the case with only one non-zero amplitude. For diffraction experiments, domains of stripe order is virtually indistinguishable from biaxial structures. Further more, the amplitudes are doubling across both a biaxial to stripe order transition and the detwinning of stripe order. The detwinning or biaxial to stripe order transitions are captured by a simple Ginzburg-Landau Hamiltonian:
\begin{align}\nonumber
    \mathcal{H}&=\kappa_1(\vert\partial_x\Phi_x\vert^2+\vert\partial_y\Phi_y\vert^2)+\kappa_2(\vert\partial_y\Phi_x\vert^2+\vert\partial_x\Phi_y\vert^2)\\\nonumber
    &+\alpha(\vert\Phi_x\vert^2+\vert\Phi_y\vert^2)+\frac{\beta}{2}(\vert\Phi_x\vert^2+\vert\Phi_y\vert^2)^2-\gamma\vert\Phi_x\vert^2\vert\Phi_y\vert^2\\\nonumber
      &   -\Delta (\vert\Phi_x\vert^2- \vert\Phi_y\vert^2)
\end{align}
where $\kappa_1, \kappa_2, \alpha, \beta$, $\gamma$ and $\Delta$ are phenomenological parameters~\cite{RobertsonPRB2006,MaestroPRB2006}. A sign change in $\alpha$ models the temperature-driven charge order symmetry breaking. Assuming $\Delta=0$, a biaxial to stripe transition appears upon sign change of $\gamma$. Only weak uniaxial pressure dependence on $\gamma$ is expected. A biaxial to stripe order transition therefore requires $\gamma\ll\beta$ to explain our observation. In this limit, higher order terms are important~\cite{RobertsonPRB2006} and hence $\Delta\neq0$. We therefore define $\Delta=\eta(\bf{r})+\epsilon$ where $\eta(\bf{r})$ takes small negative or positive values to model long range strain disorder and $\epsilon$ is a constant proportional to the applied strain. For $\alpha,\gamma<0$ and $\epsilon=0$, stripe order appears with domains as schematically illustrated in Fig.~1. External uniaxial pressure ($\epsilon>0$) lifts this domain degeneracy. We interpret the magnetic field effect inside the superconducting state as evidence for an intertwined competing interaction between these two orders. This competition can be described phenomenologically by adding the superconducting order parameter and a competing interaction terms to the Ginzburg-Landau model~\cite{ZhangPRB2002}. 

First and foremost, our study reveals charge stripe order in LSCO in its purest form. Notably, we demonstrate that stripe order that is coupled with unconventional superconductivity is an intrinsic electronic property of underdoped cuprates. In the limit of static charge stripe ordering, coupling to superconductivity is a key condition to realize an exotic PDW state\cite{AgterbergAnnRev2020}, the existence of which offers a natural explanation of the coexistence of a plethora of quantum states including superconductivity in the cuprates. Our demonstration of uniaxial-pressure-induced stripe order in the cuprate LSCO provides a clear recipe for the conditions to disentangle intertwined superconductivity and charge order and to realize high-temperature PDW order.

Beyond the cuprates, unconventional superconductivity and charge ordered states such as CDWs and electronic nematic states are coexisting or neighbouring in many material classes such as the iron pnictides~\cite{Chu2010}, cuprates~\cite{DaouNat2010,MurayamaNC2019}, ruthenates~\cite{WuPNAS2020} and heavy-fermion systems~\cite{RonningNature2017}. In the latter, the existence of a PDW has also been proposed \cite{Gerber2014}. Although the underlying microscopic details vary across these materials, the concomitant appearance of charge order and superconductivity is
universally important.
\\[1mm]

\textit{Methods}: 
High-energy (100~keV) x-ray diffraction experiments were carried at the P21.1 beamline at PETRA~III (DESY) on La$_{1.88}$Sr$_{0.12}$CuO$_4$ single crystals ($T_c=27$~K) grown by the floating zone method~\cite{ChangPRB2008}. The orthorhombic (\textit{Bmab}) structure ($T<250$~K) generates a spontaneous strain diagonally to the charge density modulations.
Our uniaxial pressure device (See Extended Data Fig.~\ref{figs1}) is compatible with both a standard displex cryostat and a 10~T cryomagnet. 
The cryomagnet sample environment restricts the accessible momentum space to the scattering plane spanned by the $a-$ and $c-$ axes. With this configuration, pressure is applied perpendicularly to the scattering plane (along the $b$-axis direction) while magnetic field points along the $c$-axis. A zero-field four-circle setup, by contrast, allows access to both the $a-c$ and $b-c$ scattering planes. In this fashion, tensile and compressive strain are accessed directly by lattice parameter measurements\cite{KimScience2018} (See Extended Data Fig.~\ref{figs2}). Uniaxial pressure along a Cu-O bond direction detwins the orthorhombic domains (see Extended Data Fig. 8 and 9). Orthorhombic domain detwinning however does not by itself influence charge order in LSCO \cite{Croft2016,Hayden2020}. Probing the charge order reflections $(\delta,0,\ell)$ and $(0,\delta,\ell)$ with $\delta\sim0.235$ and $\ell=8.5,12.5$ 
with or without an analyzer yields consistent results (See Extended Data Fig.~\ref{figs3}). Amplitude, position and width of the charge order reflections are extracted from fits to a Voigt function. Correlation lengths are defined by the inverse half-width-at-half-maximum. \\[1mm]

\textit{Authors contributions}
T.K., N.M., M.O. grew and characterised the LSCO single crystals. D. Bucher and D. Biscette designed and produced the uniaxial pressure cell. J. Choi, Q.W., S.J., N.B.C., J.K., M.H., O.I., M.v.Z., and J. Chang carried out the x-ray diffraction experiments. J. Choi, Q.W., S.J., N.B.C., J.K., and J. Chang analysed the data. M.J. and J. Chang conceived the project. J. Choi and Q. W. contributed equally. All authors contributed to the manuscript. \\[1mm]

\textit{Acknowledgements}
We acknowledge enlightening discussions with Steven Kivelson. This work was supported by the Danish Agency for Science, Technology and Innovation under DANSCATT, the Leverhulme Trust, and the Swiss National Science Foundation. We acknowledge DESY (Hamburg, Germany), a member of the Helmholtz Association HGF, for the provision of experimental facilities.
Parts of this research were carried out at PETRA III at beamline P21.1.

\bibliography{main_Nature} 

\clearpage
\onecolumngrid

\newcommand{\beginsupplement}{
        \setcounter{table}{0}
        \renewcommand{\thetable}{S\arabic{table}}
        \setcounter{figure}{0}
        \renewcommand{\figurename}{\textbf{Extended Data Figure}}}

\beginsupplement
\clearpage
\onecolumngrid

\begin{figure*}[t]
\center{\includegraphics[width=0.8\textwidth]{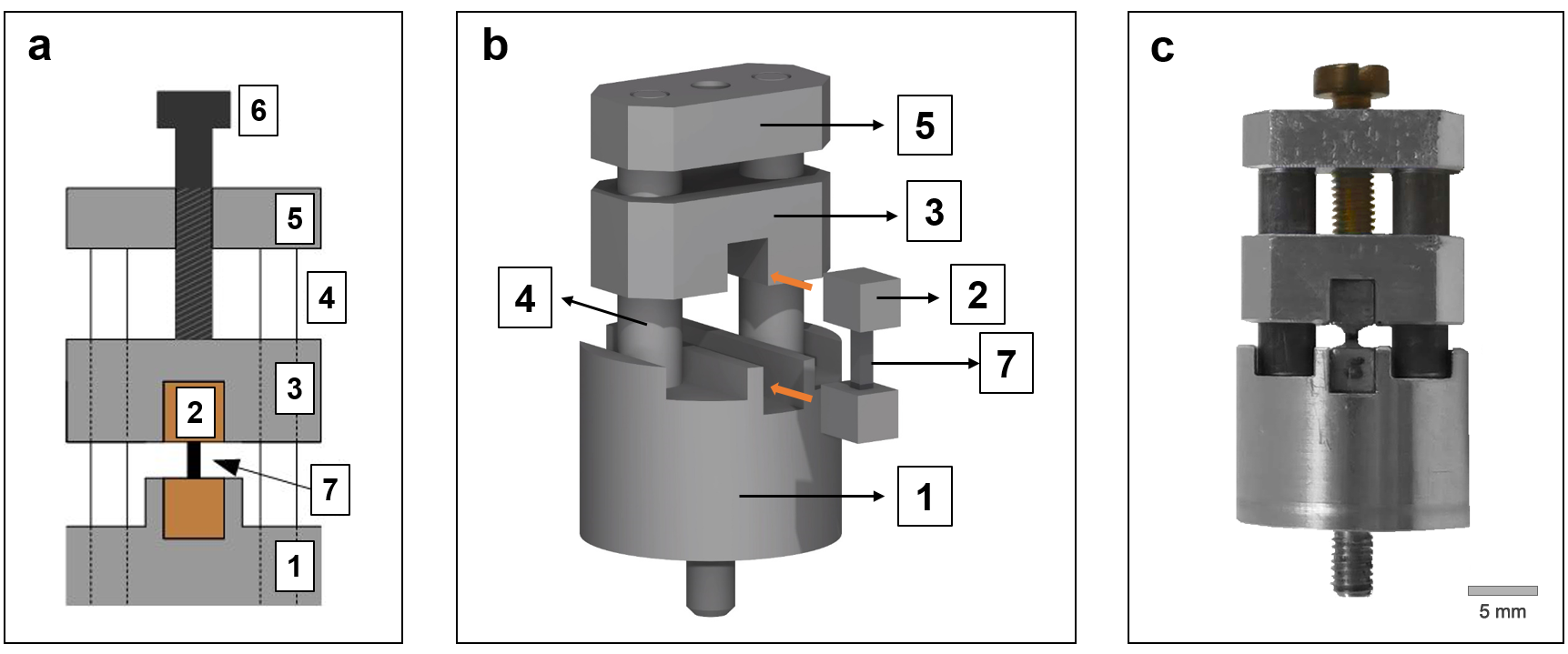}} 	
\caption{\textbf{Uniaxial pressure device for x-ray diffraction.} (\textbf{a}) Two-dimensional cross-sectional and (\textbf{b}) three-dimensional drawings of our uniaxial pressure device. 1: base plate thermally connectable to the 
cryostat, 2: sample holder, 3: middle plate, 4: guiding slide post, 5: upper plate with M3 thread, 6: M3 brass screw, 7: sample (for this study \lsco). 
Samples are glued using Stycast 2850FT or Torr Seal in between two sample holders that slides through a horizontal guided rail on the base and the middle plate, to finally reach the center position. As the top M3 screw is tightened and translated through the thread of the upper plate, the middle plate slides along the two guide posts and applies uniaxial pressure to the sample. All parts are fabricated out of non-magnetic alloys with comparable thermal contraction properties. 
(\textbf{c}) Photography of the uniaxial pressure device. The M3 screws ($\diameter=3$~mm) and the horizontal grey bar provide a metric scale. }
\label{figs1}
\end{figure*} 

\begin{figure*}[t]
\center{\includegraphics[width=1\textwidth]{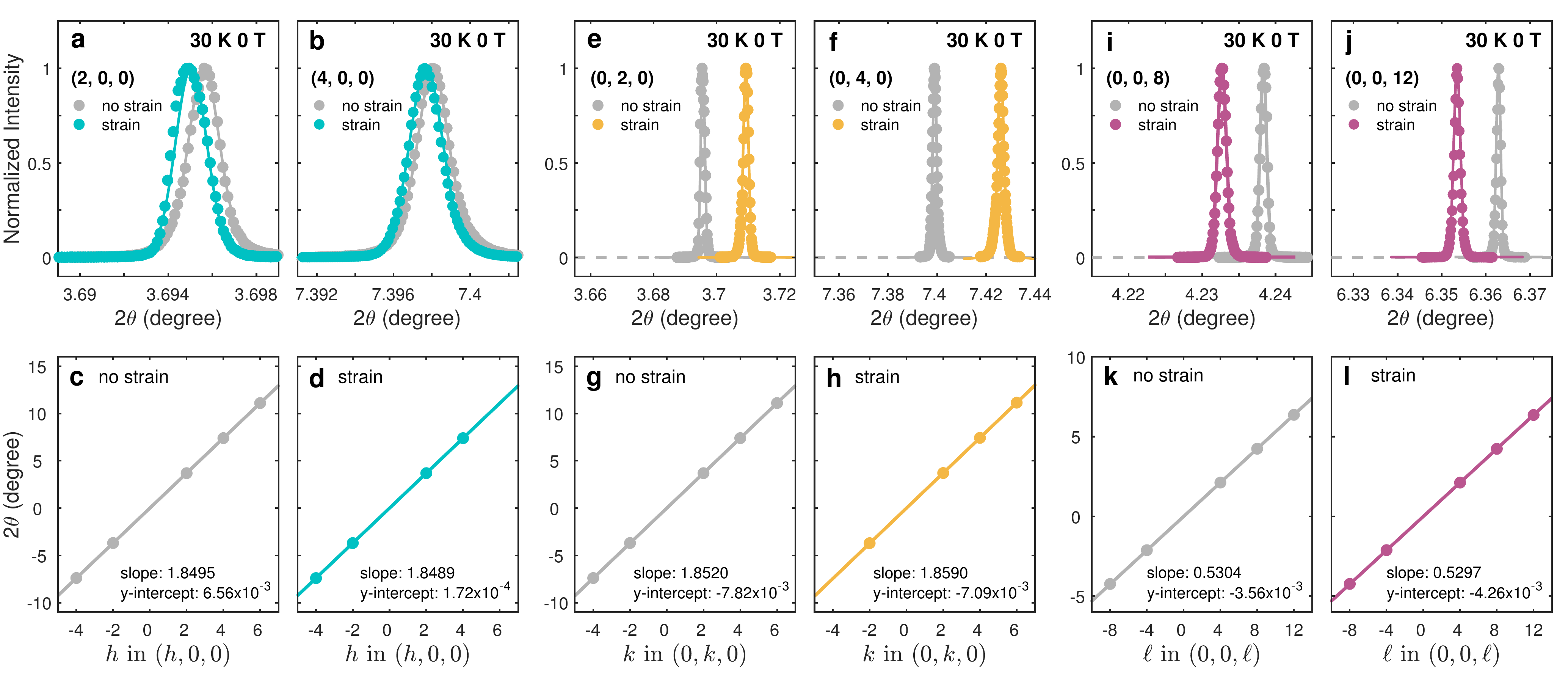}} 	
\caption{\textbf{Lattice parameter determination:} (\textbf{a,b,e,f,i,j}) Longitudinal scans through Bragg peaks $(2n,0,0)$, $(0,2n,0)$, and $(0,0,4m)$ with $n=1,2$ and $m=2,3$ for strained and unstrained conditions, respectively, with temperature and magnetic field as indicated. Solid lines are fits to a Voigt function. (\textbf{c,d,g,h,k,l}) Bragg peak positions (in $2\theta$) for $n={-2,-1,1,2,3}$ and $m=-2,-1,1,2,3$. Solid lines are linear-least-square fits permitting determination of the lattice parameters $a$, $b$ and $c$. For the unstrained condition, we find $a_0=3.7712(2)$, $b_0=3.7661(4)$, and $c_0=13.150(3)$~\AA. Strain along each direction is quantified by $\epsilon_a=(a-a_0)/a_0$, $\epsilon_b=(b-b_0)/b_0$, and $\epsilon_c=(c-c_0)/c_0$. For this particular example, we find $\epsilon_a=0.034(4)~\%$, $\epsilon_b=-0.37(2)~\%$,
and $\epsilon_c=0.13(1)~\%$. From this result, we calculate the Poisson's ratio $\nu_{ba}=0.10(1)$ and $\nu_{bc}=0.35(4)$ for LSCO. The latter ratio is in excellent agreement with what has previously been reported for LSCO~~\cite{MeyerALP2015}.
}
\label{figs2}
\end{figure*} 

\begin{figure*}[t]
\center{\includegraphics[width=0.75\textwidth]{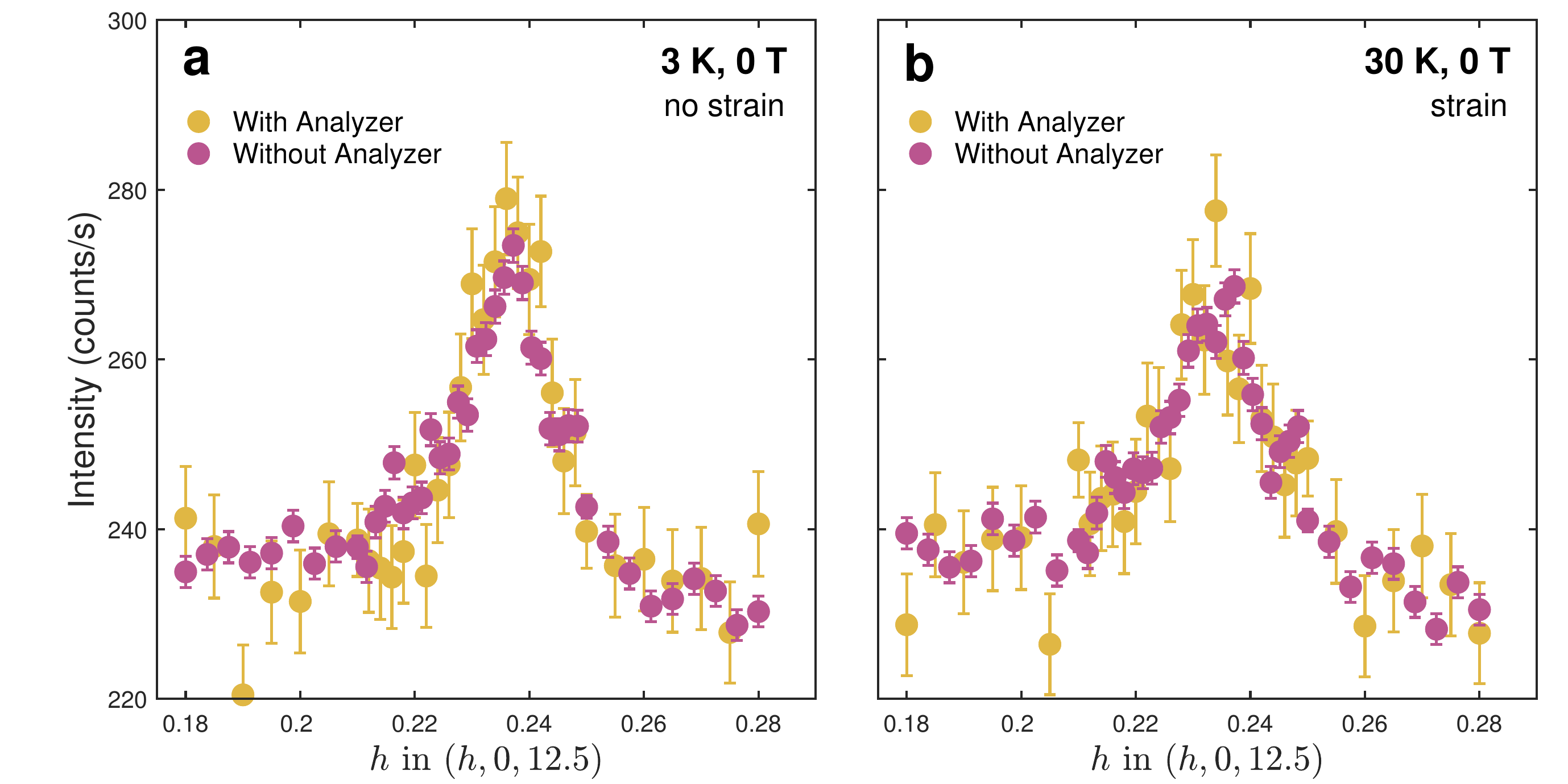}} 	
\caption{\textbf{Stripe-order reflection measured with and without an analyzer.} Longitudinal Q-scans -- through $(0.235, 0, 12.5)$ -- measured with (yellow symbols) and without an analyzer crystal (purple symbols) at (\textbf{a}) $T = 3$ K and (\textbf{b}) $T = 30$ K. Comparable profiles are found for both the triple- and two-axis measurement schemes. Data recorded without an analyzer crystal has higher counting statistics, leading to the lower error bars.}
\label{figs3}
\end{figure*} 

\newpage

\begin{figure*}[t]
\center{\includegraphics[width=0.95\textwidth]{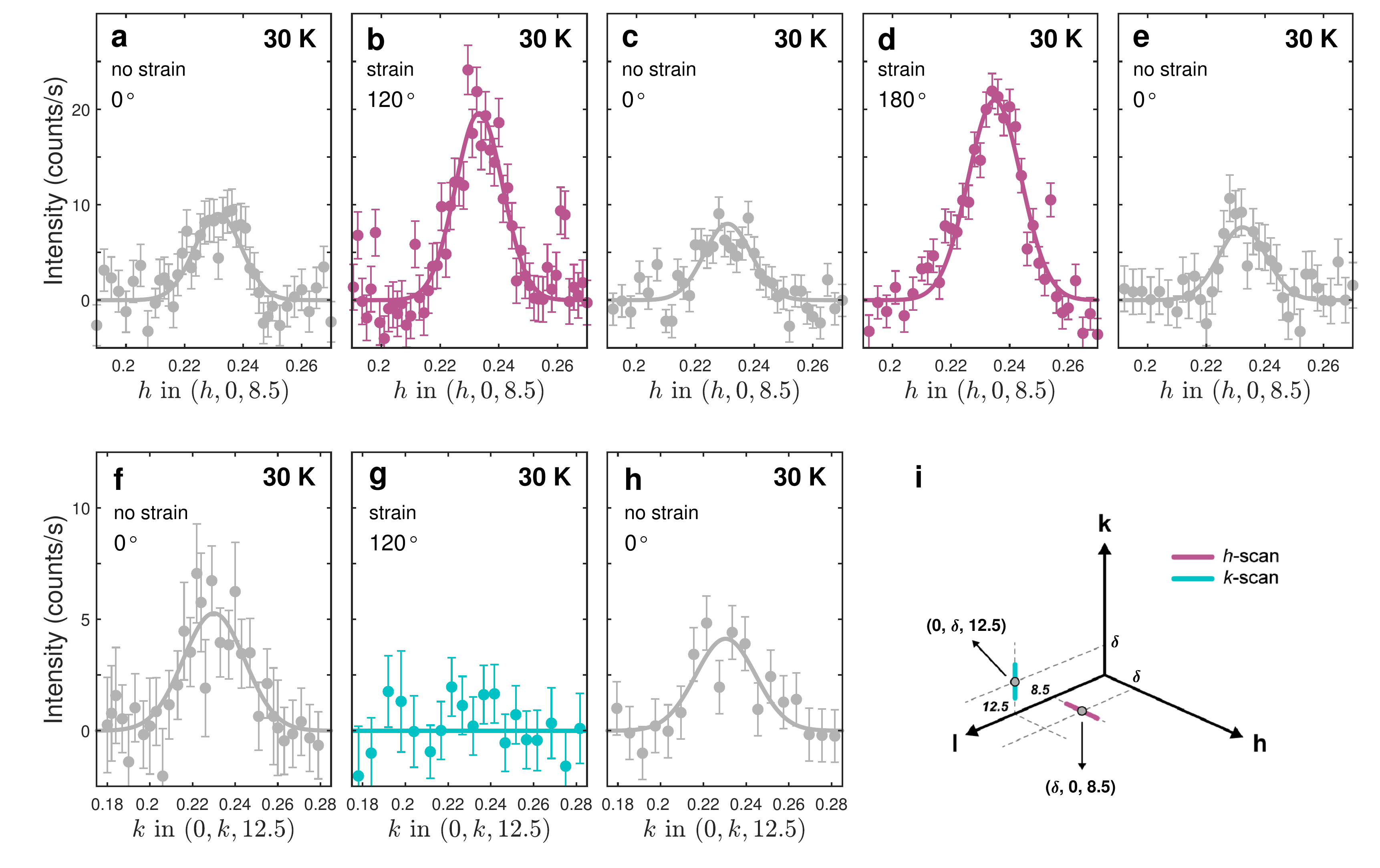}} 	
\caption{\textbf{Reproducibility and reversability of the charge-density-wave strain effect.} Intensity versus 
momentum across the charge density wave reflections (\textbf{a-e}) ($\delta$, 0, 8.5) and (\textbf{f-h}) (0, $\delta$, 12.5). These scans are measured at $T=30$~K for consecutive application  and removal 
of uniaxial pressure indicated by the screw-turn degree $\phi$. Linear  backgrounds are subtracted from the measured intensity profiles that are fitted by a Voigt function. Error bars are given by counting statistics. These results demonstrate that the uniaxial pressure effect on the charge density wave is both reproducible and reversible.}
\label{figs4}
\end{figure*} 

\begin{figure*}[t]
\center{\includegraphics[width=0.8\textwidth]{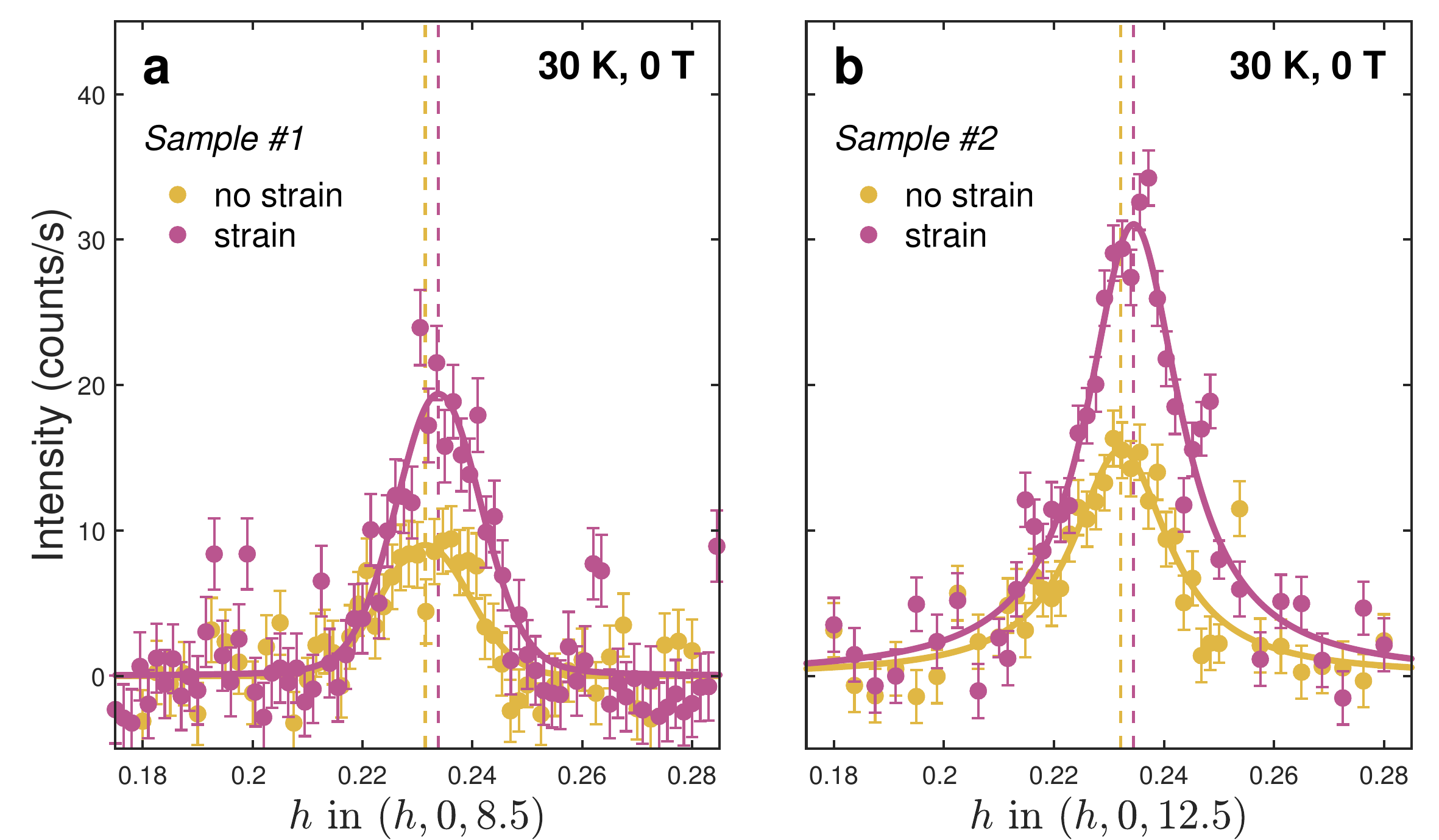}} 	
\caption{\textbf{Strain effect on two different \lsco\ single crystals.} (\textbf{a,b}) Normal state ($T=30$~ K) zero-magnetic field 
scans through respectively ($\delta$,0,8.5) and ($\delta$,0,12.5) for strained and unstrained conditions.  Two different samples (labelled $\#1$ and $\#2$) are glued using two different epoxies (Stycast 2850FT or Torr Seal). In both cases, strain induces an approximately twofold increase of the charge order diffraction intensity. The relative difference in intensity stems from different crystal size that sets the diffraction probability and the x-ray attenuation factor. Solid lines are fits to a Voigt profile whereas the vertical dashed lines indicate the fitted peak position. Error bars are set by counting statistics.}
\label{figs5}
\end{figure*}

\begin{figure*}[t]
\center{\includegraphics[width=0.8\textwidth]{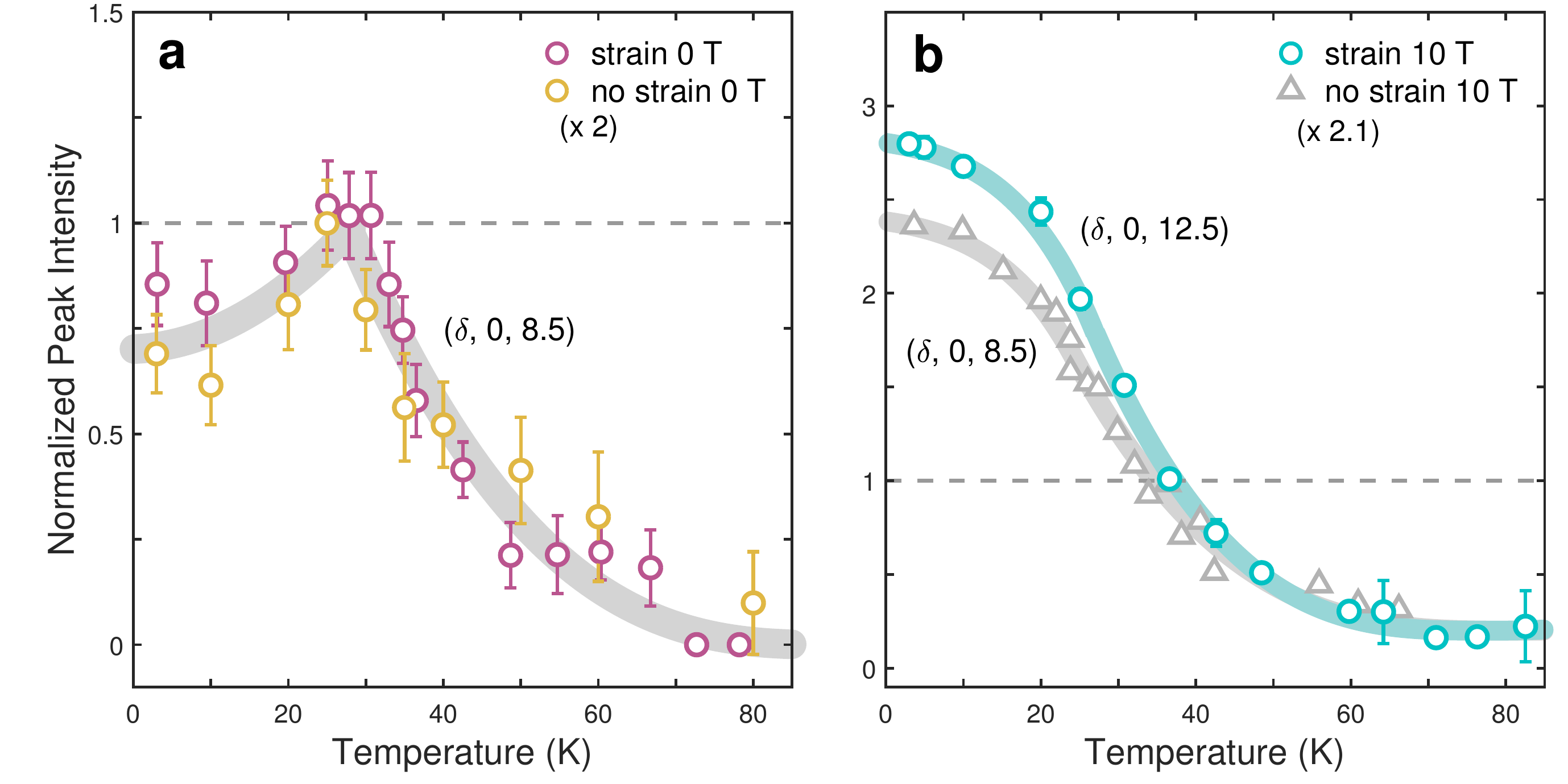}} 	
\caption{\textbf{Charge-density-wave diffraction intensity versus temperature.} (\textbf{a}) Normalized CDW intensity recorded in zero magnetic field with and without uniaxial pressure versus temperature. Both curves display a maximum at $T\approx T_c$ and a partial suppression for $T<T_c$. With the average 5~K temperature step between data points, small uniaxial pressure enhancements of the transition temperature~\cite{SASAGAWAphysicaB2005} is not expected to manifest a significant difference in this context. Within the counting statistics, the two temperature curves are identical. (\textbf{b}) CDW intensity at ($\delta$, 0, 8.5) and ($\delta$, 0, 12.5) recorded in a 10~T $c$-axis magnetic field for strained and unstrained conditions, respectively. The field effect inside the superconducting state is larger for the strained case. }
\label{figs6}
\end{figure*} 

\begin{figure*}[t]
\center{\includegraphics[width=0.75\textwidth]{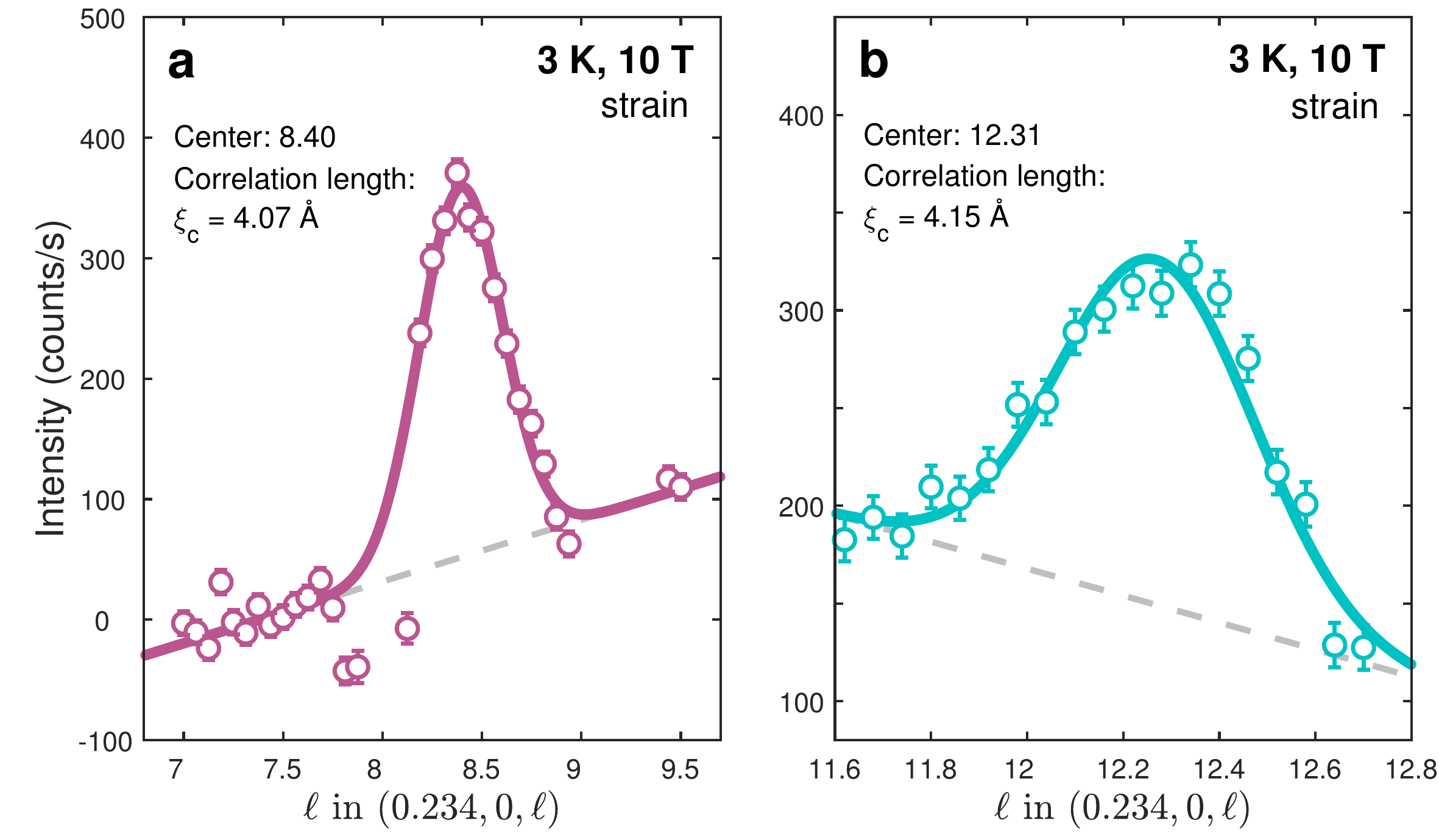}} 	
\caption{\textbf{Out-of-plane stripe-order modulation in \lsco} 
Out-of-plane $\ell$-scan through (\textbf{a}) $(\delta,0,8.5)$ and (\textbf{b}) $(\delta,0,12.5)$ for magnetic field and temperatures as indicated. The correlation length $\xi_c$ is defined by the inverse of half-width-at-half-maximum obtained from a Gaussian fit with a linear background.
}
\label{figs7}
\end{figure*} 

\begin{figure*}[t]
\center{\includegraphics[width=0.65\textwidth]{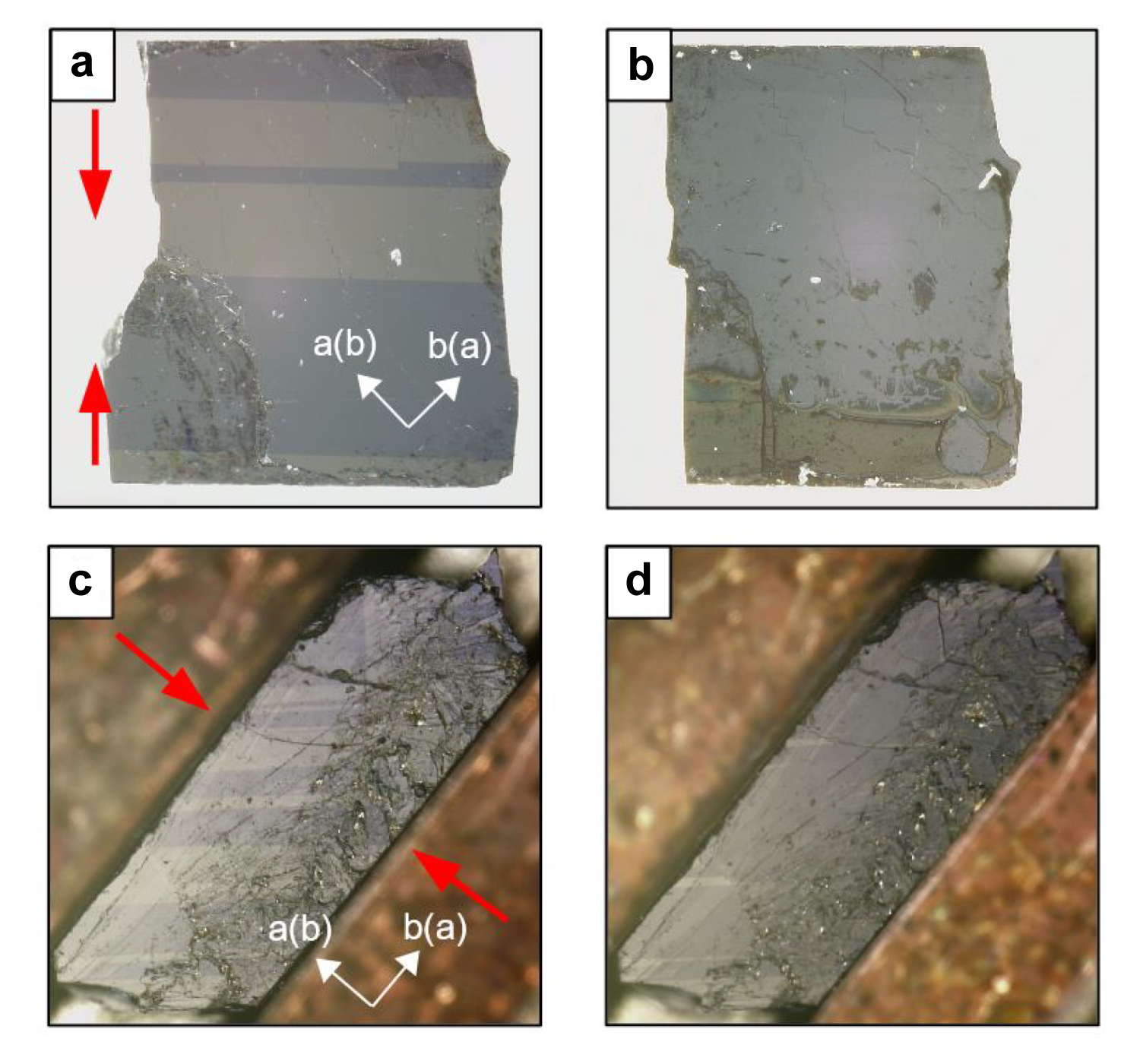}} 	
\caption{\textbf{Detwinning independent of pressure direction.} 
It is known that orthorhombic crystals can be detwinned by application of strain along the orthorhombic crystal axes. This is illustrated for Ca$_3$Ru$_2$O$_7$ in (\textbf{a,b}). The population of different orthorhombic domains is observed using a polarized light microscope. Pressing parallel or perpendicular to the domain boundaries leads to detwinning. In panels (\textbf{c,d}), we show that uniaxial pressure applied along the Ru-O bond direction (45$^{\circ}$ to the orthorhombic domain boundaries) also detwins the crystal. The direction of applied uniaxial pressure is therefore not important for the detwinning process. }
\label{figs8}
\end{figure*} 

\begin{figure*}[t]
\center{\includegraphics[width=0.75\textwidth]{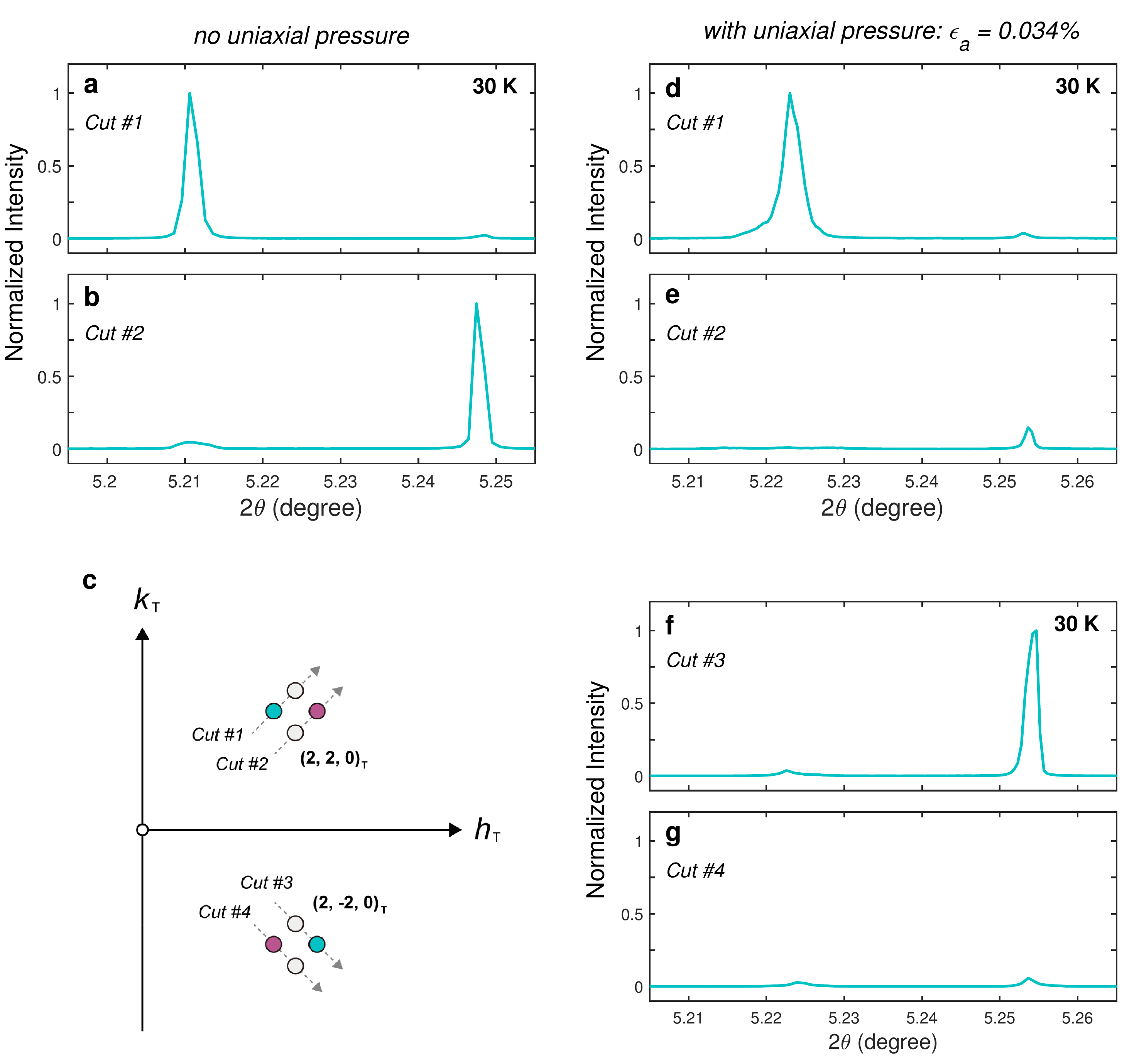}} 	
\caption{\textbf{Detwinning of \lsco.} (\textbf{a,b,d,e}) Scans through the quartet of (2,2,0) Bragg reflections originating from the orthorhombic domains - See panel (\textbf{c}).
Without uniaxial pressure, two domains dominate the population (a,b). Upon application the crystal is being almost fully detwinned as reveal only a single dominant (2,2,0) Bragg reflection (See panels \textbf{d,e}). (\textbf{f,g}) Scans through (2,-2,0) lend further support to the conclusion that uniaxial pressure along the tetragonal $b$-axis detwins the crystal.} 
\label{figs9}
\end{figure*}

\end{document}